\documentclass[reprint,aps,prl,twocolumn,superscriptaddress,showpacs,preprintnumbers,amsmath,amssymb,floatfix]{revtex4-1}

\usepackage{graphicx}
\usepackage[mathlines]{lineno}
\usepackage{xspace} 
\usepackage{wasysym}

\begin{document}
\title{First direct limits on Lightly Ionizing Particles with electric charge less than $e/6$} 


\author{R.~Agnese} \affiliation{Department of Physics, University of Florida, Gainesville, FL 32611, USA}
\author{A.J.~Anderson} \affiliation{Department of Physics, Massachusetts Institute of Technology, Cambridge, MA 02139, USA}
\author{D.~Balakishiyeva} \affiliation{Department of Physics, University of Florida, Gainesville, FL 32611, USA}
\author{R.~Basu~Thakur~} \affiliation{Fermi National Accelerator Laboratory, Batavia, IL 60510, USA}\affiliation{Department of Physics, University of Illinois at Urbana-Champaign, Urbana, IL 61801, USA}
\author{D.A.~Bauer} \affiliation{Fermi National Accelerator Laboratory, Batavia, IL 60510, USA}
\author{J.~Billard} \affiliation{Department of Physics, Massachusetts Institute of Technology, Cambridge, MA 02139, USA}
\author{A.~Borgland} \affiliation{SLAC National Accelerator Laboratory/Kavli Institute for Particle Astrophysics and Cosmology, 2575 Sand Hill Road, Menlo Park 94025, CA}
\author{M.A.~Bowles} \affiliation{Department of Physics, Syracuse University, Syracuse, NY 13244, USA}
\author{D.~Brandt} \affiliation{SLAC National Accelerator Laboratory/Kavli Institute for Particle Astrophysics and Cosmology, 2575 Sand Hill Road, Menlo Park 94025, CA}
\author{P.L.~Brink} \affiliation{SLAC National Accelerator Laboratory/Kavli Institute for Particle Astrophysics and Cosmology, 2575 Sand Hill Road, Menlo Park 94025, CA}
\author{R.~Bunker} \affiliation{Department of Physics, Syracuse University, Syracuse, NY 13244, USA}
\author{B.~Cabrera} \affiliation{Department of Physics, Stanford University, Stanford, CA 94305, USA}
\author{D.O.~Caldwell} \affiliation{Department of Physics, University of California, Santa Barbara, CA 93106, USA}
\author{D.G.~Cerdeno} \affiliation{Departamento de F\'{\i}sica Te\'orica and Instituto de F\'{\i}sica Te\'orica UAM/CSIC, Universidad Aut\'onoma de Madrid, 28049 Madrid, Spain}
\author{H.~Chagani} \affiliation{School of Physics \& Astronomy, University of Minnesota, Minneapolis, MN 55455, USA}
\author{Y.~Chen} \affiliation{Department of Physics, Syracuse University, Syracuse, NY 13244, USA}
\author{J.~Cooley} \affiliation{Department of Physics, Southern Methodist University, Dallas, TX 75275, USA}
\author{B.~Cornell} \affiliation{Division of Physics, Mathematics, \& Astronomy, California Institute of Technology, Pasadena, CA 91125, USA}
\author{C.H.~Crewdson} \affiliation{Department of Physics, Queen's University, Kingston ON, Canada K7L 3N6}
\author{P.~Cushman} \affiliation{School of Physics \& Astronomy, University of Minnesota, Minneapolis, MN 55455, USA}
\author{M.~Daal} \affiliation{Department of Physics, University of California, Berkeley, CA 94720, USA}
\author{P.C.F.~Di~Stefano} \affiliation{Department of Physics, Queen's University, Kingston ON, Canada K7L 3N6}
\author{T.~Doughty} \affiliation{Department of Physics, University of California, Berkeley, CA 94720, USA}
\author{L.~Esteban} \affiliation{Departamento de F\'{\i}sica Te\'orica and Instituto de F\'{\i}sica Te\'orica UAM/CSIC, Universidad Aut\'onoma de Madrid, 28049 Madrid, Spain}
\author{S.~Fallows} \affiliation{School of Physics \& Astronomy, University of Minnesota, Minneapolis, MN 55455, USA}
\author{E.~Figueroa-Feliciano} \affiliation{Department of Physics, Massachusetts Institute of Technology, Cambridge, MA 02139, USA}
\author{G.L.~Godfrey} \affiliation{SLAC National Accelerator Laboratory/Kavli Institute for Particle Astrophysics and Cosmology, 2575 Sand Hill Road, Menlo Park 94025, CA}
\author{S.R.~Golwala} \affiliation{Division of Physics, Mathematics, \& Astronomy, California Institute of Technology, Pasadena, CA 91125, USA}
\author{J.~Hall} \affiliation{Pacific Northwest National Laboratory, Richland, WA 99352, USA}
\author{H.R.~Harris} \affiliation{Department of Physics \& Astronomy, Texas A\&M University, College Station, TX 77843, USA}
\author{S.A.~Hertel} \affiliation{Department of Physics, Massachusetts Institute of Technology, Cambridge, MA 02139, USA}
\author{T.~Hofer} \affiliation{School of Physics \& Astronomy, University of Minnesota, Minneapolis, MN 55455, USA}
\author{D.~Holmgren} \affiliation{Fermi National Accelerator Laboratory, Batavia, IL 60510, USA}
\author{L.~Hsu} \affiliation{Fermi National Accelerator Laboratory, Batavia, IL 60510, USA}
\author{M.E.~Huber} \affiliation{Department of Physics, University of Colorado Denver, Denver, CO 80217, USA}
\author{A.~Jastram} \affiliation{Department of Physics \& Astronomy, Texas A\&M University, College Station, TX 77843, USA}
\author{O.~Kamaev} \affiliation{Department of Physics, Queen's University, Kingston ON, Canada K7L 3N6}
\author{B.~Kara} \affiliation{Department of Physics, Southern Methodist University, Dallas, TX 75275, USA}
\author{M.H.~Kelsey} \affiliation{SLAC National Accelerator Laboratory/Kavli Institute for Particle Astrophysics and Cosmology, 2575 Sand Hill Road, Menlo Park 94025, CA}
\author{A.~Kennedy} \affiliation{School of Physics \& Astronomy, University of Minnesota, Minneapolis, MN 55455, USA}
\author{M.~Kiveni} \affiliation{Department of Physics, Syracuse University, Syracuse, NY 13244, USA}
\author{K.~Koch} \affiliation{School of Physics \& Astronomy, University of Minnesota, Minneapolis, MN 55455, USA}
\author{A.~Leder} \affiliation{Department of Physics, Massachusetts Institute of Technology, Cambridge, MA 02139, USA}
\author{B.~Loer} \affiliation{Fermi National Accelerator Laboratory, Batavia, IL 60510, USA}
\author{E.~Lopez~Asamar} \affiliation{Departamento de F\'{\i}sica Te\'orica and Instituto de F\'{\i}sica Te\'orica UAM/CSIC, Universidad Aut\'onoma de Madrid, 28049 Madrid, Spain}
\author{R.~Mahapatra} \affiliation{Department of Physics \& Astronomy, Texas A\&M University, College Station, TX 77843, USA}
\author{V.~Mandic} \affiliation{School of Physics \& Astronomy, University of Minnesota, Minneapolis, MN 55455, USA}
\author{C.~Martinez} \affiliation{Department of Physics, Queen's University, Kingston ON, Canada K7L 3N6}
\author{K.A.~McCarthy} \affiliation{Department of Physics, Massachusetts Institute of Technology, Cambridge, MA 02139, USA}
\author{N.~Mirabolfathi} \affiliation{Department of Physics, University of California, Berkeley, CA 94720, USA}
\author{R.A.~Moffatt} \affiliation{Department of Physics, Stanford University, Stanford, CA 94305, USA}
\author{D.C.~Moore} \affiliation{Division of Physics, Mathematics, \& Astronomy, California Institute of Technology, Pasadena, CA 91125, USA}
\author{H.~Nelson} \affiliation{Department of Physics, University of California, Santa Barbara, CA 93106, USA}
\author{R.H.~Nelson} \affiliation{Division of Physics, Mathematics, \& Astronomy, California Institute of Technology, Pasadena, CA 91125, USA}
\author{R.W.~Ogburn} \affiliation{SLAC National Accelerator Laboratory/Kavli Institute for Particle Astrophysics and Cosmology, 2575 Sand Hill Road, Menlo Park 94025, CA} 
\author{K.~Page} \affiliation{Department of Physics, Queen's University, Kingston ON, Canada K7L 3N6}
\author{W.A.~Page} \affiliation{Department of Physics \& Astronomy, University of British Columbia, Vancouver, BC V6T 1Z1, Canada}
\author{R.~Partridge} \affiliation{SLAC National Accelerator Laboratory/Kavli Institute for Particle Astrophysics and Cosmology, 2575 Sand Hill Road, Menlo Park 94025, CA}
\author{M.~Pepin} \affiliation{School of Physics \& Astronomy, University of Minnesota, Minneapolis, MN 55455, USA}
\author{A.~Phipps} \affiliation{Department of Physics, University of California, Berkeley, CA 94720, USA}
\author{K.~Prasad} \affiliation{Department of Physics \& Astronomy, Texas A\&M University, College Station, TX 77843, USA}
\author{M.~Pyle} \affiliation{Department of Physics, University of California, Berkeley, CA 94720, USA}
\author{H.~Qiu} \affiliation{Department of Physics, Southern Methodist University, Dallas, TX 75275, USA}
\author{W.~Rau} \affiliation{Department of Physics, Queen's University, Kingston ON, Canada K7L 3N6}
\author{P.~Redl} \affiliation{Department of Physics, Stanford University, Stanford, CA 94305, USA}
\author{A.~Reisetter} \affiliation{Department of Physics, University of Evansville, Evansville, IN 47722, USA}
\author{Y.~Ricci} \affiliation{Department of Physics, Queen's University, Kingston ON, Canada K7L 3N6}
\author{H.~E.~Rogers} \affiliation{School of Physics \& Astronomy, University of Minnesota, Minneapolis, MN 55455, USA}
\author{T.~Saab} \affiliation{Department of Physics, University of Florida, Gainesville, FL 32611, USA}
\author{B.~Sadoulet} \affiliation{Department of Physics, University of California, Berkeley, CA 94720, USA}\affiliation{Lawrence Berkeley National Laboratory, Berkeley, CA 94720, USA}
\author{J.~Sander} \affiliation{Department of Physics \& Astronomy, Texas A\&M University, College Station, TX 77843, USA} \affiliation{Department of Physics, University of South Dakota, Vermillion, SD 57069, USA}
\author{K.~Schneck} \affiliation{SLAC National Accelerator Laboratory/Kavli Institute for Particle Astrophysics and Cosmology, 2575 Sand Hill Road, Menlo Park 94025, CA}
\author{R.W.~Schnee} \affiliation{Department of Physics, Syracuse University, Syracuse, NY 13244, USA}
\author{S.~Scorza} \affiliation{Department of Physics, Southern Methodist University, Dallas, TX 75275, USA}
\author{B.~Serfass} \affiliation{Department of Physics, University of California, Berkeley, CA 94720, USA}
\author{B.~Shank} \affiliation{Department of Physics, Stanford University, Stanford, CA 94305, USA}
\author{D.~Speller} \affiliation{Department of Physics, University of California, Berkeley, CA 94720, USA}
\author{S.~Upadhyayula} \affiliation{Department of Physics \& Astronomy, Texas A\&M University, College Station, TX 77843, USA}
\author{A.N.~Villano} \affiliation{School of Physics \& Astronomy, University of Minnesota, Minneapolis, MN 55455, USA}
\author{B.~Welliver} \affiliation{Department of Physics, University of Florida, Gainesville, FL 32611, USA}
\author{D.H.~Wright} \affiliation{SLAC National Accelerator Laboratory/Kavli Institute for Particle Astrophysics and Cosmology, 2575 Sand Hill Road, Menlo Park 94025, CA}
\author{S.~Yellin} \affiliation{Department of Physics, Stanford University, Stanford, CA 94305, USA}
\author{J.J.~Yen} \affiliation{Department of Physics, Stanford University, Stanford, CA 94305, USA}
\author{B.A.~Young} \affiliation{Department of Physics, Santa Clara University, Santa Clara, CA 95053, USA}
\author{J.~Zhang} \affiliation{School of Physics \& Astronomy, University of Minnesota, Minneapolis, MN 55455, USA}

\collaboration{CDMS Collaboration}

\noaffiliation

\begin{abstract}
While the Standard Model of particle physics does not include free particles with fractional charge, experimental searches have not ruled out their existence. We report results from the Cryogenic Dark Matter Search (CDMS~II) experiment that give the first direct-detection limits for cosmogenically-produced relativistic particles with electric charge lower than $e$/6. A search for tracks in the six stacked detectors of each of two of the CDMS II towers found no candidates, thereby excluding new parameter space for particles with electric charges between $e$/6 and $e$/200.

\end{abstract}


\maketitle

 
\label{intro}

Fractionally-charged particles are a well-known feature of the Standard Model (SM) of particle physics; quarks and anti-quarks possess either $\pm$2/3 or $\pm$1/3 of the electron charge, $e$.  Because of the nature of the strong interaction, however, these particles are bound inside hadrons.   Free particles with smaller charges are viable in extensions of the SM with extra
$U(1)$ gauge symmetries, a common situation in string theories~\cite{Wen:1985qj,Abel:2008ai,Shiu:2013wxa}. Some of these models
feature a dark photon with a small kinetic mixing with the SM photon~\cite{Holdom:1985ag}
that can confer an effective (very small) charge to particles in a hidden sector. Scenarios with milli-charged dark matter particles can be
constructed in this way. The question remains whether such fundamental particles with tiny electric charge exist. 
Levitometer, Millikan-droplet, collider-based experiments~\cite{Perl:2009zz}, and astrophysical (``direct search") experiments~\cite{Ambrosio:2004ub, Ambrosio:2000, Mori:1991, Aglietta:1994} have searched for free particles with fractional charges.  Direct searches are particularly interesting because energetic cosmic rays may produce fractionally charged particles with masses inaccessible to collider experiments. 
Particles with fractional charge $fe$ would lose energy at a rate proportional to $f^2$, much more slowly than known minimum ionizing particles under similar conditions~\cite{Bichsel:2006}. Direct-search experiments look for the smaller interaction energies characteristic of such Lightly Ionizing Particles (LIPs)~\cite{Ambrosio:2000}.

\begin{figure}[t]
\centering
\includegraphics[width=2.6in]{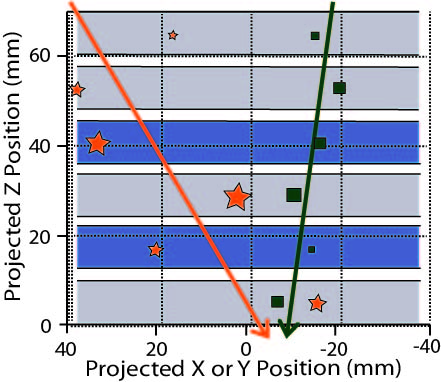}
\caption{An illustration of the difference in fit tracks through Si and Ge detectors (light and dark rectangles respectively) for a multiple-scatter Compton event and a LIP event in Tower 2. Stars (Compton event) and squares (LIP event) indicate simulated event positions; the size of the symbols is proportional to the energy deposition.}
\label{fig:hitSchematic}
\end{figure}

The Cryogenic Dark Matter Search experiment (CDMS II)~\cite{CDMSScience:2010,Akerib:2004fq}, located in the Soudan Underground Laboratory~\cite{CDMSScience:2010}, employed detectors arranged in vertical stacks of 6 detectors each (``towers"). This geometry dramatically enhances LIP background rejection when requiring all detectors along the path of a possible track to have similarly low-energy signals and hit locations that lie approximately along a straight line.  Figure~\ref{fig:hitSchematic} illustrates this rejection method by contrasting a hypothetical LIP candidate with a typical 6-detector background event from multiple Compton scattering of a photon. 
The CDMS tower geometry, coupled with the very low-energy detection thresholds ($\sim$1\,keV), provides sensitivity to LIPs with substantially smaller fractional charges than searched for in prior direct-detection experiments~\cite{Ambrosio:2004ub, Ambrosio:2000, Mori:1991, Aglietta:1994}.

The data analyzed here were collected using 5 towers of Z-sensitive Ionization and Phonon (ZIP) detectors. The detector array consisted of 19 Ge and 11 Si ZIPs, each a disk $\sim$10$\,$mm thick and 76$\,$mm in diameter. Each detector was instrumented with four phonon sensors on one face and two concentric ionization electrodes on the opposite face.  A small electric field (3\textendash4\,V/cm) was applied across the detectors to extract charge carriers created by particle interactions. The detectors were surrounded by passive lead and polyethylene shielding and cooled to $\lesssim$50$\,$mK. An outer plastic-scintillator veto identified muons with sufficient energy to penetrate the 2090 meters water-equivalent overburden of the Soudan Underground Laboratory. The CDMS II experiment was designed primarily to identify nuclear recoils from Weakly Interacting Massive Particles (WIMPs) by measuring both the ionization and phonons created by particle interactions within a detector~\cite{CDMSScience:2010}. The data described here were recorded from July 2007 to September 2008~\cite{CDMSScience:2010}. 
Only towers of fully working detectors, Tower 2 and Tower 4, were used to search for LIP tracks, resulting in raw exposures after removal of bad data periods of 59.6 and 142.4 days, respectively.

For this analysis, in which a LIP is expected to scatter from detector electrons, we require a candidate event to have a ratio of ionization to phonon energy consistent with that expected for an electron recoil. 
Candidate events are required to pass basic reconstruction-quality selection criteria similar to the criteria used in previous analyses of these data~\cite{CDMSScience:2010, CDMSLT:2011}.
In order to achieve high efficiency for LIP charges as large as $e/6$, no ionization-based fiducial-volume requirement is applied, and the energy ranges considered are 2.5--200 and 2.5--400$\,$keV for Si and Ge detectors, respectively~\cite{Prasad:2013}.  Candidate events are required to have recoil energies consistent with noise for all detectors except the six in the hit tower.


\label{analysis}

This analysis exploits the assumption that a LIP's kinetic energy is large enough that it is minimally deflected as it passes through a tower.  The resulting linear track is assumed to follow a path passing through all detectors in a tower. 
The dominant expected background is from multiply-scattering photons from the residual photon background within the shielding ($\sim$2 events keV$^{-1}$ kg$^{-1}$ day$^{-1}$)~\cite{cdms2010}. 
Figure 2 shows the results of a Geant4 Monte Carlo simulation of the photon background in which one or more consecutive detector hits are required in a single tower with no hits in any detectors of the other towers. The expected number of 6-detector-hit events in Tower 2 and Tower 4 from this simulation is 0.4$\pm$0.1 and 1.0$\pm$0.2 events, respectively.


\begin{figure}[t]
\centering
\includegraphics[width=3.2in]{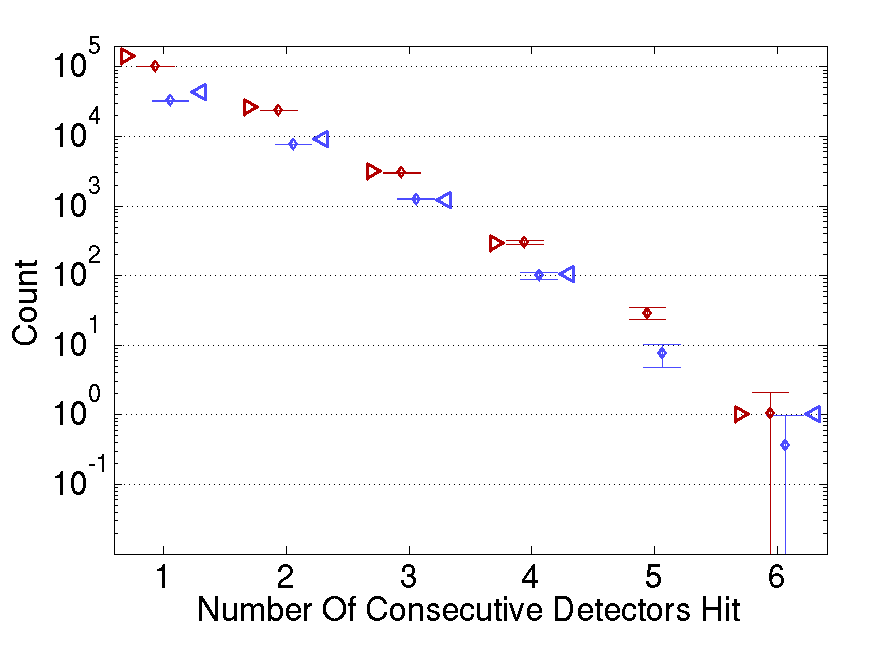}
\caption{The observed number of events, before energy-consistency and track-linearity selection criteria, as a function of the number of consecutive detectors hit, for Tower 2 ($\lhd$) and Tower 4 ($\rhd$) agrees well with the respective number of events scaled from photon background simulations ($\diamond$; performed and shown separately for each tower). The counts fall off rapidly with the number hit. Multiplicity-one and -two data are contaminated by noise and surface backgrounds~\cite{CDMSScience:2010}. Data for multiplicity-five is still blinded for a possible future analysis. Each tower simulation has a 3,150 day live time.}
\label{fig:multiplicity}
\end{figure}

\begin{figure}
\centering
\includegraphics[width=3.3in]{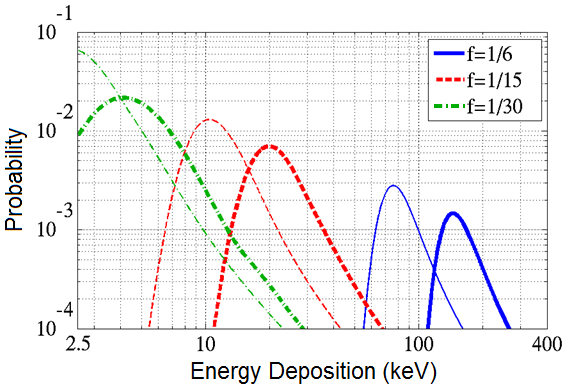}
\caption{The energy-deposition probability distributions for normally incident LIPs on 1-cm-thick Ge/Si (thick/thin curve) for three values of $f$.}
\label{fig:pdfs}
\end{figure}

A consequence of a straight track is that the distribution of observed energy depositions in the six detectors should be approximately a function (only) of $f$, the track zenith angle, and the detector material. Figure~\ref{fig:pdfs} shows the dependence of LIP energy depositions on $f$ and on the detector material for normally-incident LIPs. 
 The expected energy distributions were computed by applying the detectors' measured energy resolutions to the Photo Absorption Ionization model~\cite{Bichsel:2006}, which models the track of a relativistic LIP as a series of independent, discrete interactions with minimum ionizing energy loss (LIP $\beta\gamma\approx3$).  Any other (greater) energy loss would make the experiment more sensitive.  Assuming minimum ionizing energy loss will generally give the most ``conservative" (highest, least restrictive) upper limit in the absence of a discovery.  

\begin{figure}[t]
\centering
\includegraphics[width=3.375in]{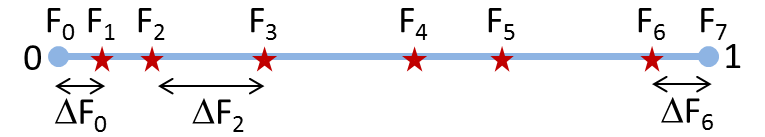}
\caption{Example of possible cumulative energy distribution values $F_i$, $i=1$ to 6, for the six detectors in a tower, showing construction of the ${\Delta}F_i$.  The $F_i$ should be approximately uniformly distributed for LIP tracks, resulting in small $E_{\mathrm{c}}$ in Equation~\ref{equationEc} (0.13 in the example shown here).} 
\label{fig:LIPCDF}
\end{figure}

\begin{figure}[t]
\centering
\includegraphics[width=3.0in]{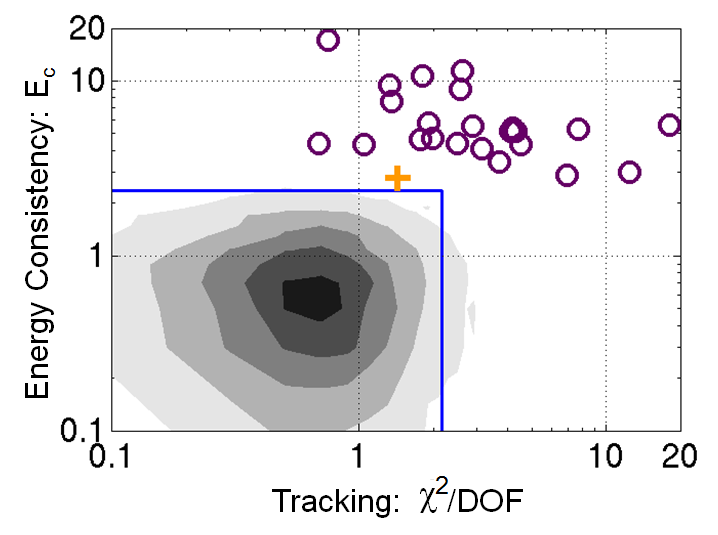}
\caption{The energy-consistency and nonblind track-linearity values of the 6-detector-hit event for Tower 4 (+) compared to simulated LIPs (10\%, 30\%, 50\% 70\%, and 90\% LIP acceptance contours in grayscale) for $f^{-1} = 15$.  No observed events or simulated 6-detector photons (o) pass both the track-linearity and energy-consistency selection criteria (solid lines). Note that the simulated photons correspond to a Tower 4 exposure that is $\sim$22$\times$ larger than in the LIPs search.}
\label{fig:trackingEnergyConsistency}
\end{figure}

We introduce a measure of energy consistency~\cite{Prasad:2013} that is small when the energy depositions in the detectors are distributed as expected and large when they are not. Let $F_i$ be the value of the cumulative probability distribution for the energy deposited in a detector, with the $F_i$ ordered from lowest to highest values for $i=1$ to 6, and include $F_0=0$ and $F_7=1$.  Define $\Delta F_i \equiv F_{i+1} - F_i$ ({\it{cf.}} Fig.~\ref{fig:LIPCDF}). The measure of energy consistency, $E_c$, is
\begin{equation}\label{equationEc}
E_c = -2\sum_{i=0}^6 w_i\ log(\Delta F_i/w_i),
\end{equation}
where the $w_i$ are weights chosen to sum to unity.  This form gives $E_c = 0$, its minimum, when each $\Delta F_i = w_i$.  The $w_i$ are chosen so that the minimum $E_c$ occurs when the $F_i$ are distributed as uniformly as possible, $w_i=1/6$ except for $w_0$ and $w_6$, which are 1/12.  The Kolmogorov-Smirnov and Cram\'er-von Mises statistics are minimized by the same distribution of energies, but this definition of $E_c$ gives a penalty that grows especially large when a detector's energy is far out on the tail of the expected distribution, as is likely for the dominant photon background.  We accept events with $E_c < 2.37$, a value chosen to have $\approx$99\% efficiency for simulated LIPs of any $f$ 
 (see Fig.~\ref{fig:trackingEnergyConsistency} for the example of $f = 1/15$).


Another consequence of minimally deflected LIPs is that the locations of each detector's interaction are expected to follow a linear track with deviations primarily due to uncertainty in position reconstruction, while background photons typically deflect from a straight track as they multiply scatter through a tower.  A full 3-dimensional $\chi^2$ fit was performed to each 6-detector track under a straight-line hypothesis.  Events were rejected when $\chi^2$/DOF $>$ 2,
approximately maximizing the expected signal-to-background ratio for all values of $f$ considered. 
In order to compute $\chi^2$, the 3-dimensional position resolution of the measurement in each detector had to be determined. 
In track fitting, the $xy$ (horizontal) coordinates are reconstructed using phonon pulse information. 
A position-independent $xy$ resolution was initially computed using phonon pulse amplitudes and varied from 5\,mm for low energy depositions to 2\,mm for high energy depositions.

The effective average $z$ (vertical) position of a detector's interactions is unknown and assumed to be at the central horizontal plane of the detector. 
The $z$-position uncertainty is a function of the assumed fractional charge and is estimated by the standard deviation of a large number of simulated events.  The simulation uses the following procedure. Whenever the sum of the energy depositions falls within the analyzed region, the simulation takes a Poisson distribution of the number of interactions, assigns each a random position and energy deposition from the single-interaction probability distribution, and computes the deviation of the energy-weighted position from where the track intersects the horizontal midplane of the detector.

All the selection criteria were set ``blind" based only on calibration data, parameter space insensitive to LIPs, and simulations, without looking at characteristics of any 6-detector events. 
Later it was realized that the $xy$-position reconstruction was significantly less accurate than originally estimated, and the resolution was worse than assumed. 
So after unblinding, we defined a new nonblind track-linearity selection criterion, but without using any of the information revealed from the unblinding.  The revised $xy$ position was estimated using not only phonon signal amplitudes, but also information from phonon relative start times and phonon pulse shapes.  We then corrected for position-dependent $xy$ biases in each detector with an algorithm~\cite{Ogburn:2008} based on the assumption that background gamma rays uniformly illuminate the detectors. After this correction, surface-to-surface events shared between adjacent detectors were used to obtain estimates of detector $xy$-position resolution. Such events typically have 2$\,$mm true lateral separation, with measured separation smeared by detector resolution.  Thus, detector resolution as a function of energy was obtained by deconvolving the observed resolution from lateral separation using conservative assumptions when necessary.  This corrected, position-dependent $xy$ resolution varies from $\sim$8--20\,mm with the worst at low energy. ``Conservative" means these estimates are expected to be larger than is realistic. Larger estimates decrease a track's computed value of $\chi^2$ and increase LIP acceptance.  In the absence of a signal, this leads to a conservative upper limit.  The improved position measurements and resolution estimates were used in the new nonblind track-linearity selection, setting the criterion 
 to maximize expected signal-to-background as a function of $f$. Monte Carlo simulations were then used to compute the LIP acceptance fraction of both the blind and nonblind track-linearity criteria.  Both blind and nonblind results are correct.  The much less sensitive blind result is also reported in order to show the effect of the original, poorly chosen, track-linearity selection criterion, with its acceptance $\sim10^{-3}$, to be compared with between 0.86 and 0.94 for the nonblind cut.
To estimate the total expected background, a scan of all values of $f^{-1}$ between 6 and 200 was performed. Figure~\ref{fig:trackingEnergyConsistency} shows the results of this scan in Tower 4 for a value of $f^{-1}$ of 15. A 6-detector-hit event may pass the track-linearity or energy-consistency value for a given $f$ range and fail for others.
This scan yielded a total expected background of
$0.16^{+0.12}_{-0.07}$  (stat) $^{+0.01}_{-0.1}$ (syst)
 events passing all nonblind selection criteria for at least one value of $f$ in the range $1/200 < f < 1/6$. 

\begin{figure}[t]
\centering
\includegraphics[width=3.0in]{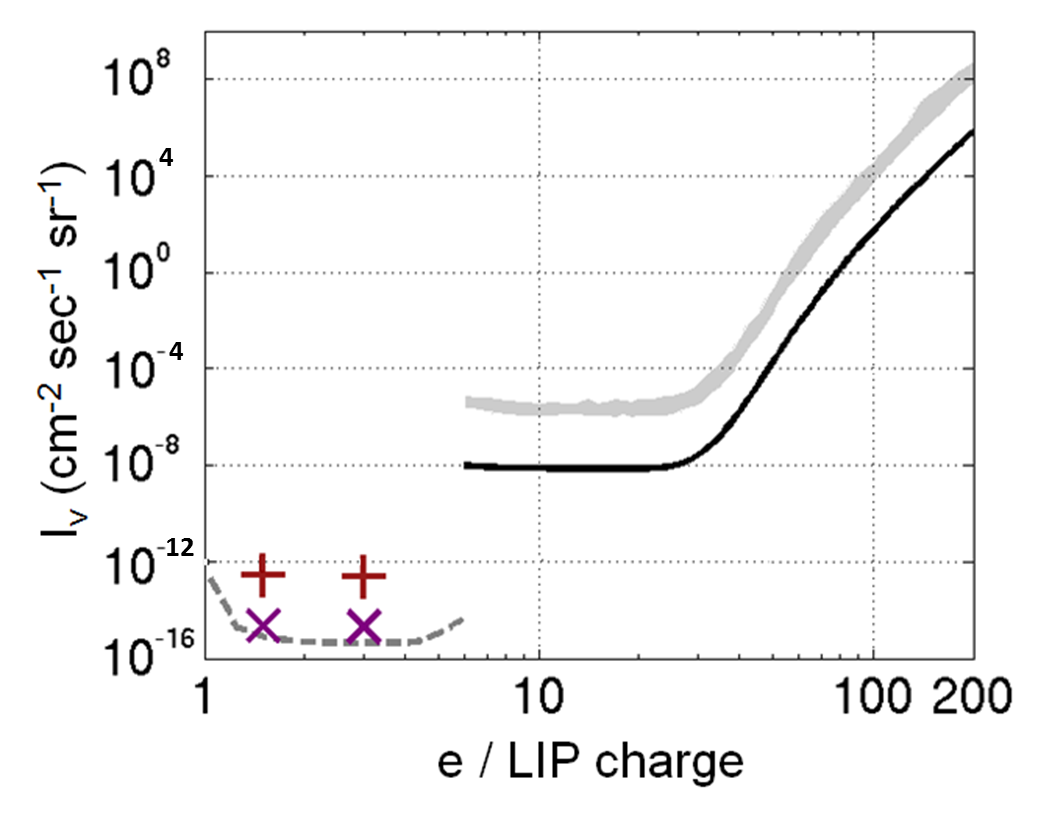}
\caption{Exclusion limits at 90\% confidence level on the LIP vertical intensity versus inverse electric charge in units of $1/e$, under the conservative assumptions that LIPs are minimum ionizing and are impinging from above, from the blind (light gray solid) and improved, nonblind (black solid) analyses of the CDMS II experiment compared with past cosmogenic searches: MACRO~\cite{Ambrosio:2004ub} (dashed lower-left), Kamiokande~\cite{Mori:1991} ($\times$), and LSD~\cite{Aglietta:1994} (+).}
\label{fig:results}
\end{figure}


The robustness of these results was probed by exploring parameter space just outside the signal region. 
Upon unblinding, but before application of the track-linearity and energy-consistency criteria, two 6-detector events (one in each tower) were observed within the analyzed region, consistent with the background prediction shown in Fig.~\ref{fig:multiplicity}.  Both events fail the blind version of the track-linearity criterion. Only one event, shown in Fig.~\ref{fig:trackingEnergyConsistency}, passes the nonblind track-linearity criterion for any value of $f$. Both also fail the energy-consistency requirement for all values of $f$. 
Figure~\ref{fig:results} shows the resulting exclusion limits on LIP vertical intensity, $I_v$, from the nonblind analysis with improved track-linearity criterion as well as from the blind analysis.  Both analyses restrict new parameter space. 

All 6-detector events narrowly outside the analyzed energy region fail at least one selection criterion. Three events in Tower 4 have an energy deposition in one detector that is marginally above the upper analysis threshold. 
Two of these near-miss events fail the energy-consistency criterion for all values of $f$ considered. The third passes the energy consistency criterion only within a narrow window of fractional charge $6<f^{-1}<9$, but it fails the track-linearity criterion for all values of $f$.

We assume LIPs are minimum ionizing. The effect of a factor $r$ increase in $dE/dx$ would be approximately cancelled by replacing $f$ by $f/\sqrt{r}$, so upper limit curves should be shifted right by a factor of $\sqrt{r}$ along the $1/f$ axis for particles with an energy loss $r$ times higher, and the limits shown in Fig.~\ref{fig:results} are conservative for all but the smallest values of $f^{-1}$.

Systematic uncertainty in LIP sensitivity is dominated by uncertainty in detector thickness (230$\,\mu$m in two Tower 4 detectors and 50$\,\mu$m otherwise), uncertainty in analysis efficiency, and accuracy of the energy calibration near thresholds.  Uncertainty in the energy calibration dominates for $f^{-1}\gtrsim30$. 
The combined systematic uncertainty for the nonblind result of less than $25\%$ is less than the width of the limit curve; the combined systematic uncertainty for the blind result is roughly given by the width of the curve. 

These results are the first to limit the flux of cosmogenic relativistic particles with charge less than $e/6$. 
Although these results are compared in Fig.~\ref{fig:results} with those from other experiments, our results were obtained at a shallower depth where the intensity is expected to be higher by a factor that depends on the LIP mass and energy spectrum. Here CDMS reports sensitivity to charges with a vertical intensity above $7\times 10^{-9}\,$cm$^{-2}\,$sec$^{-1}\,$sr$^{-1}$.  MACRO, with a substantially larger sensitive volume, achieved sensitivity down to $6.1\times 10^{-16}\,$cm$^{-2}\,$sec$^{-1}\,$sr$^{-1}$~\cite{Ambrosio:2004ub} but only for charges greater than $e/6$. The anticipated SuperCDMS SNOLAB~\cite{Sander:2012nia, Brink:2012zza} detector configuration is expected to improve sensitivity to charges  $\lesssim e/20$ because of improved energy thresholds and the use of thicker detectors.

The CDMS collaboration gratefully acknowledges the contributions of numerous 
engineers and technicians; we would like to especially thank 
Dennis Seitz, Jim Beaty, Bruce Hines, Larry Novak, 
Richard Schmitt, Astrid Tomada, and John Emes. In addition, we gratefully acknowledge assistance 
from the staff of the Soudan Underground Laboratory and the Minnesota Department of Natural Resources.
This work is supported in part by the 
National Science Foundation, by the United States Department of Energy, by NSERC Canada, and by MultiDark (Spanish MINECO). Fermilab is operated by the Fermi Research Alliance, LLC under Contract No. De-AC02-07CH11359. SLAC is operated under Contract No. DE-AC02-76SF00515 with the United States Department of Energy.


%
\bibliographystyle{apsrev4-1}

\end{document}